\documentstyle[12pt]{article}
\begin{document}


\let\al=\alpha
\let\bet=\beta
\let\ga=\gamma
\let\Ga=\Gamma
\let\de=\delta
\let\De=\Delta
\let\vareps=\varepsilon
\let\eps=\epsilon
\let\ze=\zeta
\let\ka=\kappa
\let\la=\lambda
\let\La=\Lambda
\let\ph=\varphi
\let\del=\nabla
\let\si=\sigma
\let\Si=\Sigma
\let\th=\theta
\let\tha=\vartheta
\let\Up=\Upsilon
\let\om=\omega
\let\Om=\Omega
\def\hence{~\Rightarrow}
\let\p=\partial
\let\<=\langle
\let\>=\rangle
\let\txt=\textstyle
\let\dsp=\displaystyle
\let\h=\hbox
\let\ad=\dagger
\def\e{ \hbox{e} }
\def\dot{\!\cdot\!}
\def\vfilll{\vskip 0pt plus 1filll}
\def\nl{\hfil\break}
\newcommand{\ex}[1]{\! \dot \! 10^{#1}}
\newcommand{\s}[1]{#1 \!\!/}
\def\comment#1{ \hbox{Comment suppressed here.} }
\def\tr{\mbox{Tr}\,}

%
%
%
%
\makeatletter
\def\eqnarray{\stepcounter{equation}%
              \let\@currentlabel=\theequation
              \global\@eqnswtrue
              \global\@eqcnt\z@
              \tabskip\@centering
              \let\\=\@eqncr
              $$%
 \halign to \displaywidth\bgroup
    \eqnumphantom\@eqnsel\hskip\@centering
    $\displaystyle \tabskip\z@ {##}$%
    &\global\@eqcnt\@ne \hskip 2\arraycolsep
         \hfil$\displaystyle{##}$\hfil
    &\global\@eqcnt\tw@ \hskip 2\arraycolsep
         $\displaystyle\tabskip\z@{##}$\hfil
         \tabskip\@centering
    &{##}\tabskip\z@\cr}
\def\eqnumphantom{\phantom{(\theequation)}}

\newcommand{\mbf}[1]{\mbox{\boldmath $#1$}}
\newcommand{\ra}{\rightarrow}
\newcommand{\be}{\begin{equation}}
\newcommand{\ee}{\end{equation}}
\newcommand{\ba}{\begin{eqnarray}}
\newcommand{\non}{\nonumber \\}
\newcommand{\ea}{\end{eqnarray}}
\newcommand{\baa}{\begin{eqnarray*}}
\newcommand{\eaa}{\end{eqnarray*}}
\newcommand{\barr}{\begin{array}}
\newcommand{\earr}{\end{array}}
\newcommand{\bb}{\begin{thebibliography}}
\newcommand{\eb}{\end{thebibliography}}
\newcommand{\ci}[1]{\cite{#1}}
\newcommand{\bi}[1]{\bibitem{#1}}
\newcommand{\lab}[1]{\label{#1}}
\newcommand{\re}[1]{(\ref{#1})}
\newcommand{\Tr}{\mbox{Tr\,}}
\newcommand{\tev}{{\rm ~TeV}}
\newcommand{\gev}{{\rm ~GeV}}
\newcommand{\mev}{{\rm ~MeV}}
\newcommand{\kev}{{\rm ~keV}}
\newcommand{\ev}{{\rm ~eV}}
\newcommand{\cm}{{\rm ~cm}}
\newcommand{\m}{{\rm ~m}}
\newcommand{\second}{{\rm ~sec}}
\newcommand{\SM}{the Standard Model~}
\newcommand{\PRL}[1]{~Phys.Rev.Lett.~{\bf #1}}
\newcommand{\PR}[1]{~Phys.Rev.~{\bf #1}}
\newcommand{\PL}[1]{~Phys.Lett~{\bf #1}}
\newcommand{\NP}[1]{~Nucl.Phys.~{\bf #1}}
\newcommand{\MPL}[1]{~Mod.Phys.Lett~{\bf #1}}
\newcommand{\RMP}[1]{~Rev.Mod.Phys.~{\bf #1}}
\newcommand{\etal}{{\it et al.,}~}

\let\ul=\underline
\let\ol=\overline

\def\ltap{\raisebox{-.55ex}{\rlap{$\sim$}} \raisebox{.4ex}{$<$}}
\def\gtap{\raisebox{-.55ex}{\rlap{$\sim$}} \raisebox{.4ex}{$>$}}
\def\gsim{\mathrel{\gtap}}
\def\lsim{\mathrel{\ltap}}

\hbadness=2000

\pagestyle{empty}
\begin{flushright}
{\large ITEP-M10/92 \\
July  1992}
\end{flushright}
\vspace{0.4cm}
\begin{center}
{\large \bf PAIR CORRELATOR IN THE ITZYKSON-ZUBER INTEGRAL
}\\ \vspace{1 cm} {A. Morozov}\\
\vspace{0.4 cm}
{ITEP, 117259, Moscow, Russia }
 \end{center}
\vspace{0.4cm} \noindent

\begin{abstract}
An explicit expression is suggested for the average $<U_{ij}U_{kl}^{\dagger}>$
over the uni\-ta\-ry group $SU(N)$ with the It\-zyk\-son-Zu\-ber measure
$[dU] \exp {\rm tr} \Phi U\Psi U^{\dagger}$
\end{abstract}

\section{The main result}

The remarkable fact that the integral
\ba
< 1 > \equiv \int [dU] e^{{\rm tr}\phi U\psi U^{\dagger}}
\label{IZ}
\ea
over the unitary group $SU(N)$ with the Haar measure $[dU]$ can be calculated
explicitly is extensively used in the theory of matrix models.
This integral can be
represented in terms of the eigenvalues of matrices $\phi$ and $\psi$:
\ba
< 1 > = V_N \frac{{\rm det} e^{\phi_i\psi_j}}{\Delta(\phi)\Delta(\psi)} =
c_N \sum_P (-)^P \frac{e^{\sum_k \phi_k\psi_{P(k)}}}{\Delta(\phi)\Delta(\psi)},
\label{IZ2}
\ea
where $\Delta(\phi) \equiv \prod_{i<j}^N (\phi_i-\phi_j) \equiv
\Phi_{1...N}$, $P$ are arbitrary permutations of $N$ elements and
$V_N$ is the volume of the unitary
group. This result was first obtained independently in \cite{KhCh} and
\cite{IZ}. It can be also interpreted in terms of the orbit integrals on the
lines of \cite{AFS} and in terms of the Duistermaat-Heckman (DH) theory (see
\cite{KMSW}).

In this letter the expression for the first non-trivial generalization of
\ref{IZ2} is suggested: that for the average $< U_{ij}U_{kl}^{\dagger} >$.
This quantity is of special interest for the study of observables in the
recently proposed Kazakov-Migdal model of ``induced QCD'' \cite{KM}: see
\cite{KMSW} for details. According to \cite{KMSW} in the basis where $\phi$
and $\psi$ are diagonal, the only non-vanishing pair correlators are:
\be
< \vert U_{ij}\vert^2 > \equiv \int [dU] \vert U_{ij}\vert^2
e^{\sum_{kl}\phi_k \vert U_{kl}\vert^2\psi_l} \equiv
V_N \sum_P (-)^P \frac{e^{\sum_k \phi_k\psi_{P(k)}}}{\Delta(\phi)\Delta(\psi)}
A_{ij}^{(P)}[\phi,\psi].
\label{M1}
\ee
Coefficients $A_{ij}^{(P)}$ satisfy the obvious symmetry relation
\be
A_{ij}^{(P)}[\phi,\psi] = A_{ji}^{(P^{-1})}[\psi,\phi].
\label{sym}
\ee
The answer for these quantities is most conveniently representable in the form
of a ``generating function'':
\ba
& \sum_{ij}^N \alpha_i A_{ij}^{(P)} \beta_j = &
\\
& =\sum_{n=0}^{N-1} (-)^n \sum_{i_1<...<i_{n+1}}
\frac{\sum_k (-)^k \alpha_{i_k} \Phi_{i_1...\hat{i_k}...i_{n+1}}}
{\Phi_{i_1...i_{n+1}}}
\frac{\sum_l (-)^l \beta_{P(i_l)}
\Psi_{P(i_1)...\hat{P(i_l)}...P(i_{n+1})}}
{\Psi_{P(i_1)...P(i_{n+1})}}. &
\nonumber
\label{main}
\ea
Here $\Phi_{i_1...i_p} \equiv \prod_{i_a < i_b} (\phi_{i_a}-\phi_{i_b})$,
while $\Phi$ for $p=0$ and $p=1$ are defined to be equal to unity; $\Psi$ are
defined in the same way. Hat over the index means that it is omited.

To clarify the rather sophisticated notation, it is instructive to look at
somewhat more explicit expression for the first terms with $n=0,1,2$ in
(\ref{main}):
\ba
& \sum_{i=1}^N \alpha_i\beta_{P(i)} -
\sum_{i<j}^N \frac{(\alpha_i-\alpha_j)(\beta_{P(i)}-\beta_{P(j)})}{\Phi_{ij}
\Psi_{P(i)P(j)}}~+~&  \\ & +
\sum_{i<j<k}^N \frac{(\alpha_i\Phi_{jk} - \alpha_j\Phi_{ik}+
\alpha_k\Phi_{ij})(\beta_{P(i)}\Psi_{P(j)P(k)} -
\beta_{P(j)}\Psi_{P(i)P(k)}~+~\beta_{P(k)}\Psi_{P(i)P(j)})}
{\Phi_{ijk}\Psi_{P(i)P(j)P(k)}}~+~\ldots &
\nonumber
\label{M2}
\ea
or
\ba
 & A_{ij}^{(P)} = \delta_{jP(i)} \left( 1 - \sum_{k\neq
i}\frac{1}{\phi_i-\phi_k}
\frac{1}{\psi_{P(i)}-\psi_{P(k)}}~+~\right.  &  \nonumber \\
& +
\left.
\sum_{k\neq l\neq i \; k\neq i}\frac{1}{(\phi_i-\phi_k)(\phi_i-\phi_l)}
\frac{1}{(\psi_{P(i)}-\psi_{P(k)})(\psi_{P(i)}-\psi_{P(l)})} - \ldots
\right)~+~&
\nonumber \\
& (1~-~\delta_{jP(i)}) \left( -\frac{1}{\phi_i-\phi_{P^{-1}(j)}}
\frac{1}{\psi_j - \psi_{P(i)}}~+~\right. & \\ &~+~
\left.
\sum_{l\neq P^{-1}(j)\neq i \; l\neq i} \frac{1}{(\phi_i-\phi_{P^{-1}(j)})
(\phi_i-\phi_l)} \frac{1}{(\psi_j - \psi_{P(i)})(\psi_j-\psi_{P(l)})} - \ldots
\right) &
\nonumber
\label{M3}
\ea

\section{Example of N=2}

In the case of $N=2$ we get from (\ref{main}) and (\ref{M3}):
\ba
& < \vert U_{11}\vert^2 > = < \vert U_{22} \vert^2 > =
V_2 \left( \frac{e^{\phi_1\psi_1+\phi_2\psi_2}}{\Phi_{12}\Psi_{12}}
(1~-~\frac{1}{\Phi_{12}\Psi_{12}}) ~+~ \frac{e^{\phi_1\psi_2+\phi_2\psi_1}}
{(\Phi_{12}\Psi_{12})^2} \right) &
\nonumber \\
& < \vert U_{12} \vert^2 > = < \vert U_{21}\vert^2 > =
V_2 \left( \frac{e^{\phi_1\psi_1+\phi_2\psi_2}}{(\Phi_{12}\Psi_{12})^2}
- \frac{e^{\phi_1\psi_2+\phi_2\psi_1}}{\Phi_{12}\Psi_{12}}
(1~+~\frac{1}{\Phi_{12}\Psi_{12}}) \right) &
\label{N=2}
\ea
These formulas can be easily $derived$ from the two identites:
\be
\sum_{j=1}^N < \vert U_{ij}\vert^2 > ~=~< 1 >~=
\sum_{i=1}^N < \vert U_{ij}\vert^2 >,
\label{unit}
\ee
and
\ba
& \sum_{j=1}^N < \vert U_{ij}\vert^2 >\psi_j ~=~
\frac{\partial}{\partial\phi_i}
<1> ~=~  &
\nonumber \\
& V_N \sum_P (-)^P \frac{e^{\sum_k
\phi_k\psi_{P(k)}}}{\Delta(\phi)\Delta(\psi)}
\left( \psi_{P(i)} - \sum_k\frac{1}{\phi_i -\phi_k} \right) &
\label{der}
\ea
(there is of course a similar formula for $\psi_j$-derivative of $<1>$).
These identities, while valid for any $N$, are sufficient for the non-ambiguous
evaluation of $< \vert U_{ij}\vert^2 >$ only if $N=2$. Still, one of the
arguments in favour of (\ref{main}) is that it satisfies both (\ref{unit})
and (\ref{der}), because
\be
\sum_k (-)^{k-1} \Phi_{i_1...\hat{i_k}...i_{n+1}} = \delta_{n,0}
\label{unit2}
\ee
and
\be
\sum_k (-)^{k-1} \phi_{i_k}\Phi_{i_1...\hat{i_k}...i_{n+1}} =
\phi_{i_1}\delta_{n,0}~+~\Phi_{i_1i_2}\delta_{n,1}.
\label{der2}
\ee

\section{Example of N=3}

{}From (\ref{main}) and (\ref{M3}) we obtain for $N=3$:
\ba
< \vert U_{11} \vert^2 > =\frac{V_3}{\Delta(\phi)\Delta(\psi)}
\left\{ e^{\phi_1\psi_1+\phi_2\psi_2+\phi_3\psi_3}
\left( 1 - \frac{1}{\Phi_{12}\Psi_{12}} - \frac{1}{\Phi_{13}
\Psi_{13}}+\frac{1}{\Phi_{12}\Phi_{13}\Psi_{12}\Psi_{13}}
\right) + \right. & &
\nonumber \\
{}~+~e^{\phi_1\psi_1+\phi_2\psi_3+\phi_3\psi_2}
\left( -1 ~+~ \frac{1}{\Phi_{12}\Psi_{13}}~+~\frac{1}{\Phi_{13}
\Psi_{12}} - \frac{1}{\Phi_{12}\Phi_{13}\Psi_{12}\Psi_{13}}
\right)~+~& &
\non
{}~+~e^{\phi_1\psi_2+\phi_2\psi_1+
\phi_3\psi_3}
\left( \frac{1}{\Phi_{12}\Psi_{12}} -
\frac{1}{\Phi_{12}\Phi_{13}\Psi_{12}\Psi_{13}} \right)~+~& &
\non
  ~+~ e^{\phi_1\psi_2+\phi_2\psi_3+\phi_3\psi_1}
\left( -\frac{1}{\Phi_{13}\Psi_{12}}~+~
\frac{1}{\Phi_{12}\Phi_{13}\Psi_{12}\Psi_{13}} \right) ~+~  & &
\non
 ~+~ e^{\phi_1\psi_3+\phi_2\psi_2+\phi_3\psi_1}
\left( \frac{1}{\Phi_{13}\Psi_{13}} -
\frac{1}{\Phi_{12}\Phi_{13}\Psi_{12}\Psi_{13}} \right) ~+~ & &
\non
 \left.
{}~+~ e^{\phi_1\psi_3+\phi_2\psi_1+\phi_3\psi_2}
\left( -\frac{1}{\Phi_{12}\Psi_{13}}~+~
\frac{1}{\Phi_{12}\Phi_{13}\Psi_{12}\Psi_{13}} \right)
\right\}  & &
\label{N=3.1}
\ea
\ba
 < \vert U_{12}\vert^2 >~=~\frac{V_3}{\Delta(\phi)\Delta(\psi)}
\left\{
e^{\phi_1\psi_1+\phi_2\psi_2+\phi_3\psi_3}
\left( \frac{1}{\Phi_{12}\Psi_{12}} -
\frac{1}{\Phi_{12}\Phi_{13}\Psi_{12}\Psi_{23}} \right) ~+~
\right.  & &
\non
+ e^{\phi_1\psi_1+\phi_2\psi_3+\phi_3\psi_2}
\left( -\frac{1}{\Phi_{13}\Psi_{12}}~+~
\frac{1}{\Phi_{12}\Phi_{13}\Psi_{12}\Psi_{23}} \right) ~+~ & &
\non
+e^{\phi_1\psi_2+\phi_2\psi_1+\phi_3\psi_3}
\left( -1 ~-~\frac{1}{\Phi_{12}\Psi_{12}}~+~\frac{1}{\Phi_{13}
\Psi_{23}}~+~\frac{1}{\Phi_{12}\Phi_{13}\Psi_{12}\Psi_{23}}
\right) ~+~ & &
\non
+ e^{\phi_1\psi_2+\phi_2\psi_3+\phi_3\psi_1}
\left( 1 ~-~ \frac{1}{\Phi_{12}\Psi_{23}}~+~\frac{1}{\Phi_{13}
\Psi_{12}} - \frac{1}{\Phi_{12}\Phi_{13}\Psi_{13}\Psi_{23}}
\right) ~+~ & &
\non
+e^{\phi_1\psi_3+\phi_2\psi_2+\phi_3\psi_1}
\left(\frac{1}{\Phi_{12}\Psi_{23}}~+~
\frac{1}{\Phi_{12}\Phi_{13}\Psi_{12}\Psi_{23}}\right) ~+~ & &
\non
 \left.
+e^{\phi_1\psi_3+\phi_2\psi_1+\phi_3\psi_2}
\left( -\frac{1}{\Phi_{13}
\Phi_{23}} - \frac{1}{\Phi_{12}\Phi_{13}\Psi_{12}\Psi_{23}}
\right) \right\} & &
\label{N=3.2}
\ea
\ba
 < \vert U_{13} \vert^2 >~=~\frac{V_3}{\Delta(\phi)\Delta(\psi)}
\left\{
e^{\phi_1\psi_1+\phi_2\psi_2+\phi_3\psi_3}
\left( \frac{1}{\Phi_{13}\Psi_{13}}~+~
\frac{1}{\Phi_{12}\Phi_{13}\Psi_{13}\Psi_{23}} \right) ~+~
\right. & &
\nonumber \\
+e^{\phi_1\psi_1+\phi_2\psi_3+\phi_3\psi_2}
\left( -\frac{1}{\Phi_{12}\Psi_{13}} -
\frac{1}{\Phi_{12}\Phi_{13}\Psi_{13}\Psi_{23}} \right) ~+~& &
\non
+e^{\phi_1\psi_2+\phi_2\psi_1+\phi_2\psi_3}
\left( -\frac{1}{\Phi_{13}\Psi_{23}} -
\frac{1}{\Phi_{12}\Phi_{13}\Psi_{13}\Psi_{23}} \right) ~+~& &
\non
+e^{\phi_1\psi_2+\phi_2\psi_3+\phi_3\psi_1}
\left( \frac{1}{\Phi_{12}\Psi_{23}} +
\frac{1}{\Phi_{12}\Phi_{13}\Psi_{13}\Psi_{23}} \right) ~+~
 & &
\nonumber \\
+e^{\phi_1\psi_3+\phi_2\psi_2+\phi_3\psi_1}
\left( -1~-~\frac{1}{\Phi_{12}\Psi_{23}} -
\frac{1}{\Phi_{13}\Psi_{13}} -
\frac{1}{\Phi_{12}\Phi_{13}\Psi_{13}\Psi_{23}} \right) ~+~ & &
\non
\left.
+ e^{\phi_1\psi_3+\phi_2\psi_1+\phi_3\psi_2}
\left( 1~+~\frac{1}{\Phi_{12}\Psi_{13}} +
\frac{1}{\Phi_{13}\Psi_{23}}~+~
\frac{1}{\Phi_{12}\Phi_{13}\Psi_{13}\Psi_{23}} \right)
\right\}. & &
\label{N=3.3}
\ea
Other $< \vert U_{ij}\vert^2 >$ are given by similar expressions.

In order to found these formulas in the case of $N=3$ it is enough to
supplement (\ref{unit}) and (\ref{der}) by the requirement that $< \vert
U_{ij}\vert^2 >$ remain finite whenever any pair of $\phi-$ or $\psi-$
eigenvalues coincide.
In fact it appears sufficient to consider only two cases:
$\psi_2 = \psi_3$ and $\phi_1 = \phi_2$.
After (\ref{N=3.1}-\ref{N=3.3}) are derived, eq.(\ref{main}) arises as a
straightforward generalization for the case of arbitrary $N$ (the form of
this expression is again very much restricted by the finiteness condition).

\section{Conclusion and Acknowledgements}

Straightforward derivation can be obtained either in the orbit-
or in the DH-like
tecknique. I know from S.Shatashvili that he in fact found an explicit
derivation of
$< U_{ij}U_{kl}^{\dagger} >$ in the orbit approach \cite{S},
giving it in the
elegant form of an integral over Gelfand-Tseytlin parameters, which
seems to reproduce (\ref{main}) after integration.
Representation of this kind should be especially convenient for the
study of the large-$N$ limit.
(I understand also that A.Migdal found a
formula in the large-$N$ approximation by a more straightforward method.)
Representation like (\ref{main})
should be especially convenient for analysis in the DH-framework, as
suggested in \cite{KMSW}.
Anyhow, it is now clear that the formulas
for pair correlators appear to be
rather simple,  and this supports the suggestion of ref.\cite{KMSW}
that all the
correlators with Itzykson-Zuber measure can be exactly calculable
 (and this in fact can be done by the method of \cite{S}, though the
problem of evaluation of the generating functional
$< \exp A_{ij}\vert U_{ij} \vert^2 >$ remains to be resolved).

I am indebted to I.Kogan, A.Niemi, G.Semenoff, S.Shatashvili and N.Weiss
for stimulating discussions. I also acknowledge the hospitality of the
Physical Departement of UBC where the main part of this work was done and
the Mathematical Departement of Columbia University where it was
completed.

\end{document}